\documentclass[lettersize,journal]{IEEEtran}
\IEEEoverridecommandlockouts

\usepackage{cite}
\usepackage[subrefformat=parens]{subcaption}   
\usepackage{tabularx}     
\usepackage{textcomp}     
\usepackage{threeparttable} 
\usepackage{nth}          
\usepackage{gensymb}      
\usepackage{graphicx}     
\usepackage{graphics}
\usepackage{mathtools}    
\usepackage{amsmath}
\usepackage{amsfonts}
\usepackage{amsthm}
\usepackage{amssymb}
\usepackage{threeparttable}

\usepackage[usenames, dvipsnames]{color}

\usepackage{setspace}
\usepackage{caption}
\usepackage{todonotes}
\usepackage{bbm}

\usepackage[linesnumbered,vlined,ruled]{algorithm2e}

\newcommand\scalemath[2]{\scalebox{#1}{\mbox{\ensuremath{\displaystyle #2}}}}

\newcommand{\EE}{\mathbb{E}}
\newcommand{\one}{\mathbbm{1}}

\usepackage{tikz}

\usepackage{xcolor}

\allowdisplaybreaks[1]

\begin{document}

\title{Autocorrelation and Spectrum Analysis for Variable Symbol Length Communications with Feedback}


\author{ Chin-Wei Hsu, Hun-Seok Kim,~\IEEEmembership{Senior Member,~IEEE},
Achilleas Anastasopoulos,~\IEEEmembership{Senior Member,~IEEE} 
\thanks{The authors are with the Department of Electrical Engineering and Computer Science, University of Michigan, Ann Arbor, MI, 48109, USA. (e-mail: chinweih@umich.edu; hunseok@umich.edu; anastas@umich.edu)
}
}


\IEEEaftertitletext{\vspace{-2\baselineskip}}

\maketitle

\begin{abstract}
Variable-length feedback codes can provide advantages over fixed-length feedback or non-feedback codes.
This letter focuses on uncoded variable-symbol-length feedback communication and analyzes the autocorrelation and spectrum of the signal.
We provide a mathematical expression for the autocorrelation that can be evaluated numerically. 
We then numerically evaluate the autocorrelation and spectrum for the variable-symbol-length signal in a feedback-based communication system that attains a target reliability for every symbol by adapting the symbol length to the noise realization.
The analysis and numerical results show that the spectrum changes with SNR when the average symbol length is fixed, and approaches the fixed-length scheme at high SNR.
\end{abstract}



\IEEEpeerreviewmaketitle


\vspace{-3mm}
\section{Introduction}

Feedback in communications enables variable-length codes.
Compared to fixed-length codes, variable-length codes have advantages such as better error exponent \cite{Polyanskiy2011} or tradeoff between error probability and transmission time in fading channels \cite{Lott2007}.
Two examples of variable-length codes are the Yamamoto-Itoh scheme~\cite{YI1979} and the Burnashev scheme~\cite{Burnashev1976}.
In the Yamamoto-Itoh scheme, the decoder requests block retransmissions through feedback depending on its confidence on the last block, and hence the number of blocks (also the total number of symbols) is variable.
In the Burnashev scheme, a posterior probability mass function (pmf) of the message is kept and updated on both encoder and decoder for each channel usage, and the next symbol to transmit as well as the number of transmitted symbols are determined by this pmf. The number of symbols is variable depending on the noise realization.
In both cases, the length of a single symbol in the block is a constant, while the variation in the block length comes from the variable number of symbols transmitted.
Comparison with fixed-length codes is done under the constraint that the expected block length matches that of a fixed-length scheme.

Instead of having variable number of symbols in a block with fixed symbol length, one can envision a variable-length code with a fixed number of symbols where each symbol may have a variable length.
This idea dates back to Viterbi's work in 1965 \cite{Viterbi1965}, where he considered a continuous antipodal signal that is only terminated by a feedback signal when the decoder observes a certain reliability of the symbol.
Recently, a novel feedback code scheme based on this idea, called opportunistic symbol length adaptation (OSLA), was proposed in~\cite{OSLA_globecom}, and it was shown that it outperforms state-of-the-art practical feedback codes in both noiseless and noisy feedback scenarios.

To evaluate the error rate performance of different schemes under a fair comparison, signal-to-noise-ratio (SNR) and spectral efficiency are required to be the same.
For schemes with constant symbol length, it is easy to equate spectral efficiency by simply choosing same symbol length and code rate.
However, this is not straightforward for variable symbol length schemes without evaluating the spectrum of the signal.

This letter analyzes the autocorrelation and spectrum of a variable-symbol-length signal in communications with feedback.
We derive from first principles the autocorrelation function given an arbitrary probability density function (pdf) for the symbol length.
We provide an expression for the autocorrelation function that can be easily evaluated through numerical integration, and can be used to evaluate the spectrum through numerical evaluation of the corresponding Fourier transform.
Finally, we numerically evaluate the autocorrelation and spectrum of the OSLA signal, and show that the change in the occupied bandwidth compared to a fixed symbol length scheme is negligible when the expected symbol length is matched.
Therefore, we claim that the error rate performance evaluation in~\cite{OSLA_globecom} is done in a fair way compared to a fixed-length scheme and other variable-length schemes.

\vspace{-2mm}
\section{Motivating Communication Scenario}

To motivate this work, we consider an uncoded feedback communication system
that was first proposed in~\cite{Viterbi1965} and it is further 
generalized to the OSLA system in~\cite{OSLA_globecom}.
Without any coding, the system can attain 6~dB SNR gain for the same error probability compared to a conventional fixed-symbol-length scheme when the expected symbol lengths are matched~\cite{Viterbi1965}.
When channel coding is applied, with the help of a more sophisticated decoding algorithm, this scheme provides an additional gain of up to 1.5 dB without any change in the code structure \cite{OSLA_globecom}.
Moreover, the system shows superior feedback robustness and block length scalability compared to other state-of-the-art feedback codes~\cite{deepcode2020, Ben2017}.

\begin{figure}[ht]
\includegraphics[width=\linewidth]{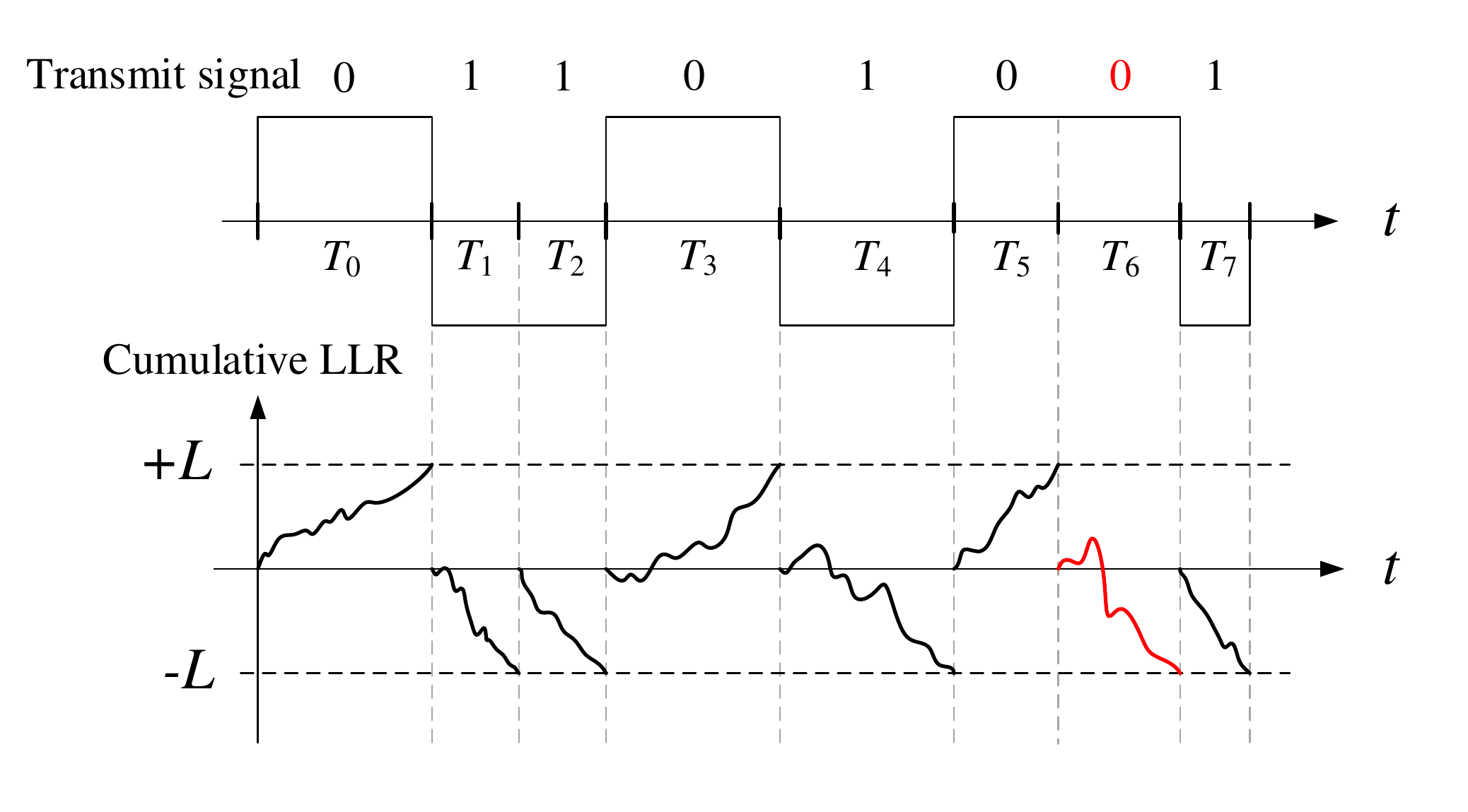}
\centering
\caption{OSLA waveform and cumulative LLR. Observe that there is an erroneous decision for bit~6.}
\label{fig:OSLA}
\end{figure}

The system is composed of a source and a destination, connected through a forward channel and a feedback channel. 
The source continuously transmits an antipodal symbol via the forward channel until the destination sends a termination signal via the feedback channel.
The destination continuously calculates the cumulative log-likelihood ratio (LLR) of the antipodal symbol until its absolute value reaches a predetermined threshold $L$. 
Once this happens, the destination sends the termination signal via the noiseless feedback channel and the transmission of the next antipodal signal starts.
The resulting transmitted signal consists of antipodal rectangular pulses, each having a random duration depending on the noise realization.
This transmission process is illustrated in Figure~\ref{fig:OSLA}, where the symbol length is determined by the timing of the cumulative LLR reaching $+L$ or $-L$.
A symbol is demodulated incorrectly when the cumulative LLR reaches the opposite boundary.

\vspace{-5mm}

Specifically, for the transmission of $K+1$ symbols the transmitted signal can be expressed as 
\begin{align} \label{eq:signal}
X(t) = \sum_{k=0}^{K} B_k \  p\Big(\frac{t-S_k}{T_k}\Big),
\end{align}
where $B_k$'s are i.i.d. random variables (r.v.) with equal probability of being $+1$ or $-1$, and 
\begin{align}
p(t) = 
	\begin{cases}
	1 &,0 < t < 1\\
	0 &,\text{otherwise}.
	\end{cases}
\end{align}
With this expression, the pulse corresponding to the $k$-th symbol has duration equal to $T_k$ and the starting time of the pulse is $S_k$, where $S_k$ is defined as
\begin{align}\label{eq:Sdef}
S_k = 
	\begin{cases}
	0 &, k=0 \\
	\sum_{i=0}^{k-1} T_i &, k \ge 1.
	\end{cases}
\end{align}

To make the discussion more concrete, we now consider the statistics of the i.i.d random variables $T_k$ for the system described in~\cite{Viterbi1965,OSLA_globecom}.
The LLR value calculated at the destination can be considered as a Wiener process with nonzero drift~\cite{WienerBook}. 
The symbol length $T$ is a stopping time, defined as the first time the LLR reaches the boundary $+L$ or $-L$.
The pdf of $T$ in this case is given by \cite[page 233]{WienerBook}
\begin{align}
f_T(t) = \frac{Le^{-\gamma t}(e^{-\frac{L}{2}} + e^{\frac{L}{2}})}{\sqrt{16\pi\gamma t^3}} \sum_{k=-\infty}^\infty (1+4k)e^{-\frac{L^2(1+4k)^2}{16\gamma t}},
\label{eq:pdf_exact}
\end{align}
or in an approximation form when $L\gg 1$ \cite[page 223]{WienerBook}
\begin{align}
f_T(t) \approx \frac{L}{\sqrt{16\pi\gamma t^3}} e^{-\frac{(L-4\gamma t)^2}{16\gamma t}},
\label{eq:T_pdf}
\end{align}
where $\gamma=\frac{P}{N_0}$, $P$ is the power of the signal and $N_0$ is the noise spectral density level.
The average energy per bit is given by $E_b=P \mathbb{E}\{T\}$.

\section{Power Spectral Density Evaluation}

The autocorrelation function of the signal in~\eqref{eq:signal} is defined as
\vspace{-2mm}
\begin{align}
R(t,\tau) = \mathbb{E}\{ X(t+\tau)X(t) \}.
\end{align}
In the following, we only consider the case $\tau > 0$. The case of $\tau \leq 0$ can be derived similarly.
It is easy to see that
\begin{align}
R(t,\tau) &= \sum_{k=0}^K \mathbb{E}\big\{ p\Big(\frac{t+\tau-S_k}{T_k}\Big) p\Big(\frac{t-S_k}{T_k}\Big) \big\} \nonumber \\
&= \sum_{k=0}^K A_k = A_0 + \sum_{k=1}^K A_k
\end{align}
where $A_k$ denotes each of the expectation terms inside the summation.
In the following, anticipating the fact that we will consider $R(t,\tau)$ for $t\rightarrow \infty$ we only consider these expressions for $t>0$.

First consider the term $A_0$:
\begin{align}
A_0 &= \EE \{ p\Big(\frac{t+\tau}{T_0}\Big) p\Big( \frac{t}{T_0}\Big)\} \nonumber \\
&= \EE \{ \one(0 < t < T_0-\tau) \} 
= 1 - F_T (t+\tau), 
\label{eq:A_0}
\end{align}
where $\one(\cdot)$ is the indicator function such that it equals to 1 if the condition in the parentheses is satisfied, and 0 otherwise, and $F_T(\cdot)$ is the cdf of $T$.

Next, consider $A_k$ for $k \ge 1$. 
Note that due to~\eqref{eq:Sdef}, $T_k$ and $S_k$ are independent random variables. As a result we have
\begin{align}
A_k &= \EE_{T_k} \EE_{S_k} \{ p\Big( \frac{t+\tau-S_k}{T_k}\Big) p\Big( \frac{t-S_k}{T_k}\Big) \} \nonumber \\
&= \EE_{T_k} \Big\{ \int_{-\infty}^\infty f_{S_k}(s) p\Big( \frac{t+\tau-s}{T_k}\Big) p\Big( \frac{t-s}{T_k}\Big) ds \Big\} \nonumber \\
&= \EE_{T_k} \Big\{  \int_{t+\tau-T_k}^{t} f_{S_k}(s) \one(\tau < T_k) ds \Big\} \nonumber \\
&= \EE_T \Big\{ \one(\tau < T) \big[ F_{S_k}(t) - F_{S_k}(t+\tau-T) \big] \Big\}.
\label{eq:A_k_plus}
\end{align}

Combining these two expressions, we have
\begin{align}
R(t, &\tau) =  1-F_T(t+\tau) \nonumber \\
&+ \EE_T \Big\{  \one(T>\tau) \sum_{k = 1}^K \big[ F_{S_k}(t) - F_{S_k}(t+\tau-T) \big].
\label{eq:R_4term}
\end{align}


\vspace{-1mm}
The expression in \eqref{eq:R_4term} requires the evaluation of $F_{S_k}(t)$. This can be done as follows.
\begin{align}
F_{S_k}(t) &= \mathbb{P} (S_k \le t) \nonumber \\
&= \int_{0}^\infty \mathbb{P}( S_k \le t| T_k=x)f_T(x) dx \nonumber \\
&= \int_{0}^\infty \mathbb{P}( S_{k-1} \le t - x| T_k=x)f_T(x) dx \nonumber \\
&= \int_{0}^\infty \mathbb{P}( S_{k-1} \le t - x)f_T(x) dx \nonumber \\
&= \int_{0}^\infty F_{S_{k-1}}(t-x) f_T(x) dx \nonumber \\
&= F_{S_{k-1}}(t) \ast f_T(t) \nonumber \\
&= \cdots \nonumber \nonumber \\
&= F_{S_{1}}(t) \ast f_T(t) \ast \cdots \ast f_T(t),
\end{align}
where $\ast$ denotes convolution.
Observe that
\begin{align}
 F_{S_{1}}(t) = F_T(t) = \int_0^t f_T(x) dx = u(t) \ast f_T(t)
\end{align}
where $u(t)$ is the step function.
Therefore, $F_{S_k}(t)$ can be obtained by convolving $u(t)$ with $f_T(t)$ $k$ times.

It is easier to express these convolutions in frequency domain, where convolution can be replaced by multiplication.
Let $\Phi_k(\omega), \Phi(\omega)$ and $U(\omega)$ be the Fourier transform of $F_{S_k}(t), f_T(t)$ and $u(t)$, respectively\footnote{Although the Fourier Transform of $u(t)$ does not exist, we use it formally in our expressions. As can be seen from the final result in~\eqref{eq:R_positive_t}, the expressions depend on Fourier transforms that are well defined.}.
Then
\begin{align}
\Phi_k(\omega) = U(\omega) \Phi^k(\omega).
\end{align} 
Similarly, to obtain $F_{S_k}(t+\tau-T)$, we can use its Fourier transform $\Phi_k(\omega)e^{-j\omega(T-\tau)}$.

Therefore, the expectation term in~\eqref{eq:R_4term} becomes (for $\omega>0$)
\begin{align}
& \scalemath{.87}{\EE \Big\{ \one(T>\tau) \sum_{k= 1}^K \mathcal{F}^{-1}\big\{ U(\omega)\Phi^k(\omega) - U(\omega)\Phi^k(\omega)e^{-j\omega(T-\tau)}\big\} \Big\} }\nonumber \\
&= \scalemath{.95}{\EE \Big\{ \one(T>\tau) \sum_{k= 1}^K \mathcal{F}^{-1}\big\{ U(\omega)\Phi^k(\omega)(1-e^{-j\omega(T-\tau)}) \big\} \Big\} } \nonumber \\
&= \scalemath{.89}{\EE \Big\{ \one(T>\tau)  \mathcal{F}^{-1}\big\{ U(\omega) \big[\sum_{k= 1}^K \Phi^k(\omega)\big](1-e^{-j\omega(T-\tau)}) \big\} \Big\} } \nonumber \\
&= \scalemath{.87}{\EE \Big\{ \one(T>\tau)  \mathcal{F}^{-1}\big\{ U(\omega) \frac{\Phi(\omega)(1-\Phi^{K}(\omega))}{1-\Phi(\omega)} (1-e^{-j\omega(T-\tau)}) \big\} \Big\} } \nonumber \\
&= \scalemath{.87}{\mathcal{F}^{-1}\Big \{  \frac{U(\omega)\Phi(\omega)(1-\Phi^{K}(\omega))}{1-\Phi(\omega)} \EE_T \big\{ \one(T>\tau) \big[ 1-e^{-j\omega(T-\tau)} \big] \big\} \Big\} } \nonumber \\
&= \scalemath{.82}{\mathcal{F}^{-1} \Big \{  \frac{U(\omega)\Phi(\omega)(1-\Phi^{K}(\omega))}{1-\Phi(\omega)} \int_\tau^\infty f_T(s)(1-e^{j\omega(\tau-s)})ds \Big\}.}
\end{align}
Let
\vspace{-2mm}
\begin{align}
Q(\omega ; \tau) = \int_\tau^\infty f_T(s)(1-e^{j\omega(\tau-s)})ds 
\end{align}
\vspace{-2mm}
and
\begin{align} \label{eq:G}
G(\omega ; \tau) = \frac{\Phi(\omega)(1-\Phi^{K}(\omega))}{1-\Phi(\omega)} Q(\omega ; \tau), \ \ \ \omega>0.
\end{align}
Then, for $t,\tau>0$ we have
\begin{align}
R(t, \tau) &= 1-F_T(t+\tau) + \mathcal{F}^{-1} \Big\{ U(\omega) G(\omega; \tau) \Big\} \nonumber \\
&= 1-F_T(t+\tau) + u(t) \ast g(t; \tau) \nonumber \\
&= 1-F_T(t+\tau) + \int_{-\infty}^t g(s; \tau) ds
\label{eq:R_positive_t}
\end{align}
where $g(t; \tau)$ is the inverse Fourier transform of $G(\omega; \tau)$ (w.r.t the first argument).

Note that the autocorrelation function $R(t,\tau)$ depends on both $t$ and $\tau$ as expected.
Unlike the standard case for constant $T$ where the autocorrelation becomes cyclostationary and a time randomization argument is invoked to average out the dependence on $t$, here we follow a different approach.
First we consider the limit of an infinite sequence of symbols, i.e., we consider $K\rightarrow \infty$ for fixed $t,\tau$. 
Taking this limit results in exactly the same equations with the exception of~\eqref{eq:G} that simplifies\footnote{Note that $|\Phi(\omega)| = |\EE\{e^{-j\omega T}\}| < \EE\{|e^{-j\omega T}|\}=1$ for $\omega > 0$ due to Jensen's inequality and the assumption that $T$ is not a constant and has a continuous pdf. Therefore we have $\Phi^K(\omega) \rightarrow 0$ and thus the limit result.} to
$G(\omega ; \tau) = \frac{\Phi(\omega)}{1-\Phi(\omega)} Q(\omega ; \tau)$.

Furthermore, it is observed (as shown in Figure \ref{fig:R_converge}) that for $T$ is not a constant, the autocorrelation function $R(t,\tau)$ converges for large $t$.
To get rid of $t$, we can choose $t \rightarrow \infty$ when $K=\infty$, and define $R(\tau) = \lim_{t\rightarrow \infty} R(t, \tau)$.
This is a reasonable choice if we don't know where $t=0$ is and we just sample the signal at a random time.

From \eqref{eq:R_positive_t}, we have
\vspace{-2mm}
\begin{align}
\lim_{t\rightarrow \infty}R(t,\tau) &= \int_{-\infty}^{\infty} g(s;\tau) ds 
= \lim_{\omega\rightarrow 0} G(\omega,\tau) \nonumber \\
&= \lim_{\omega\rightarrow 0} \frac{\Phi(\omega)Q(\omega; \tau)}{1-\Phi(\omega)} 
= \lim_{\omega\rightarrow 0} \frac{Q(\omega; \tau)}{1-\Phi(\omega)} \nonumber \\
&= - \frac{Q'(\omega;\tau)}{\Phi'(\omega)} \Big|_{\omega=0}, 
\end{align}
where the last equality is due to L'Hospital's rule.
%
For the numerator we have
\begin{align}
Q'(\omega; \tau)\Big|_{\omega=0} &= \frac{d}{d\omega} \int_\tau^\infty f_T(s)(1-e^{j\omega(\tau-s)})ds \Big|_{\omega=0} \nonumber \\
&= \int_\tau^\infty  f_T(s) ( -j)(\tau-s) e^{j\omega(\tau-s)} ds \Big|_{\omega=0} \nonumber \\
&= -j \int_\tau^\infty   (\tau-s) f_T(s) ds.
\end{align}

For the denominator we have
\begin{align}
\Phi'(\omega)\Big|_{\omega=0} &=\frac{d}{d\omega}\int_{-\infty}^{\infty} f_T(s)e^{-j\omega s} ds \Big|_{\omega=0} \nonumber \\
&= \int_{-\infty}^\infty f_T(s)  (-js) e^{-j\omega s} ds \Big|_{\omega=0} \nonumber \\
&= -j \int_{-\infty}^\infty s f_T(s) ds 
= -j \mathbb{E}\{T\}.
\end{align}

Therefore, we have for $\tau>0$
\begin{align}
R(\tau) &= -\frac{-j \int_\tau^\infty (\tau-s) f_T(s) ds}{-j \mathbb{E}\{T\}} \nonumber \\
&= \frac{1}{\mathbb{E}\{T\}} \int_\tau^\infty (s-\tau) f_T(s) ds,
\label{eq:R_tau}
\end{align}
and due to symmetry $R(-\tau)=R(\tau)$ for $\tau>0$.

The signal spectrum $S(f)$ can be obtained by taking the Fourier transform of $R(\tau)$. 
However, without a close-form expression of $R(\tau)$, it is hard to obtain an analytical solution of $S(f)$.
Nevertheless, given a known pdf $f_T$, the spectrum can be assessed numerically.

\vspace{-2mm}
\section{Numerical Results}

We numerically evaluate $R(t,\tau)$ using the analytical results in \eqref{eq:R_positive_t}, and compare to the simulation results.
Note that the pdf in \eqref{eq:T_pdf} depends on $\gamma = P/N_0$ and $L$.
Since the analysis aims at providing a spectrum comparison with the fixed-symbol-length scheme, here we always choose a $L$ such that $\mathbb{E}\{T\}=1$ for all $\gamma$.
In this case, $E_b/N_0 = P\cdot \mathbb{E}\{T\}/N_0 = \gamma$.
In the following, we show numerical results of the analysis and the simulation by generating $T_k$ with pdf $f_T$ in \eqref{eq:pdf_exact}.
We use $K=100$ to simulate a relatively long sequence.

\begin{figure}[ht]
\includegraphics[width=0.9\linewidth]{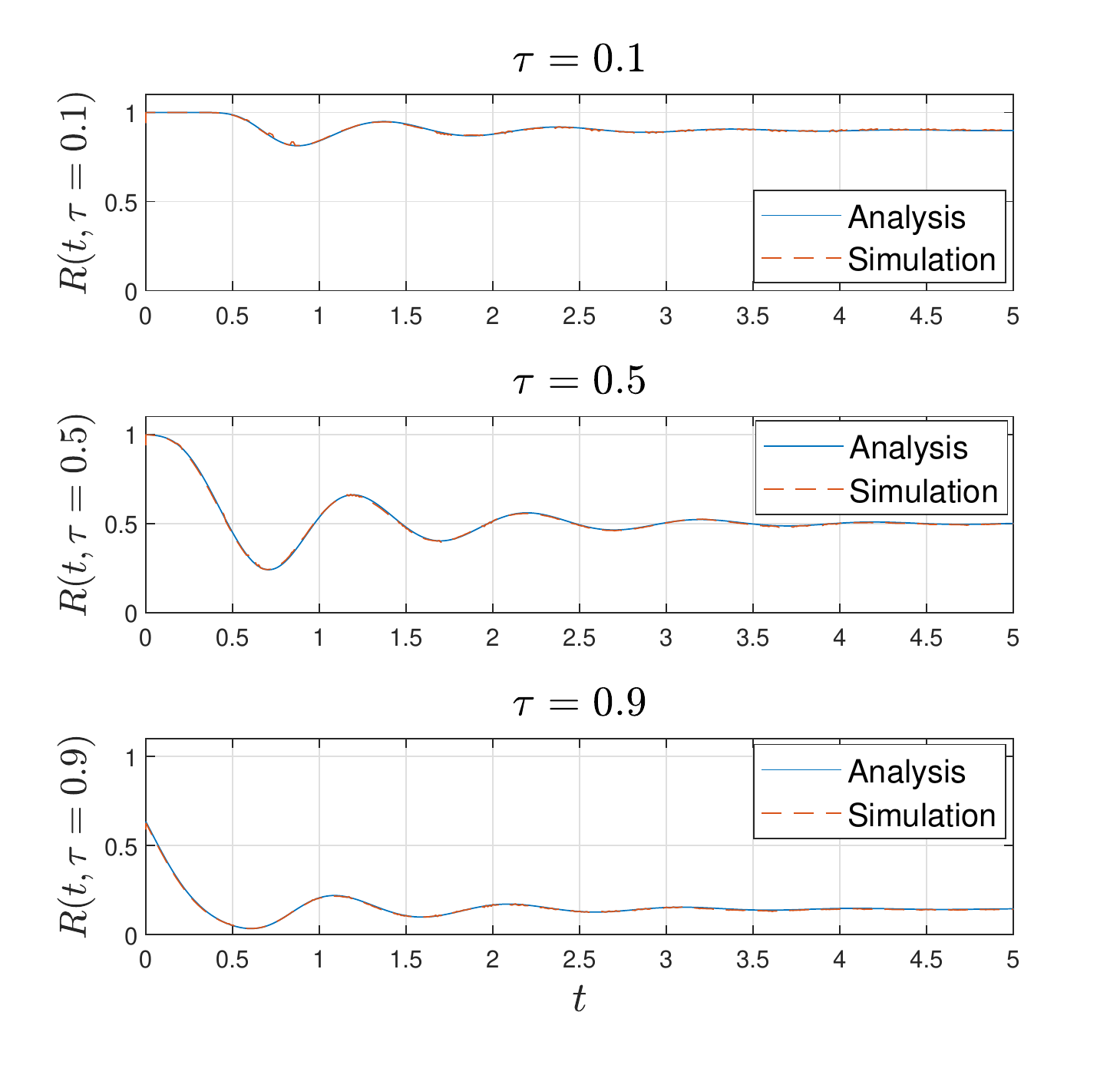}
\centering
\vspace{-4mm}
\caption{$R(t,\tau)$ analysis and simulation. $E_b/N_0=10$dB.}
\label{fig:R_converge}
\end{figure}

In Figure \ref{fig:R_converge}, it is observed that for a fixed $\tau$, the value of $R(t,\tau)$ depends on $t$, but converges to a fixed value as $t \rightarrow \infty$. 

\begin{figure}[ht]
\includegraphics[width=0.85\linewidth]{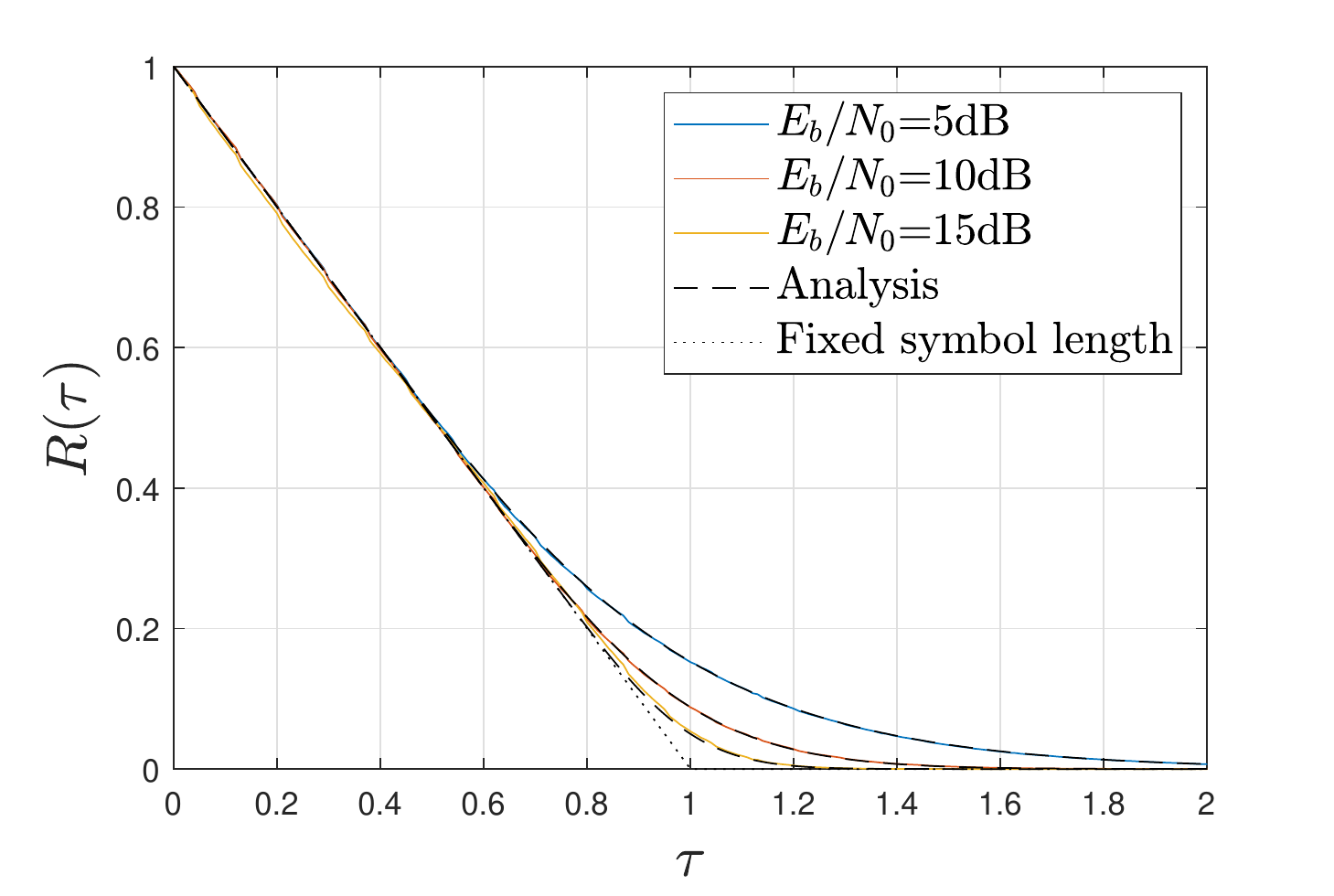}
\centering
\vspace{-2mm}
\caption{$R(\tau)$ of OSLA signal at various $E_b/N_0$.}
\label{fig:R_inf}
\end{figure}

Figure \ref{fig:R_inf} shows the analysis from \eqref{eq:R_tau} and simulation results of the autocorrelation function at different $\gamma=E_b/N_0$.
It is observed that for high SNR, the shape of $R(\tau)$ approachs a fixed-length scheme.
This is expected since at large $\gamma$, the pdf in \eqref{eq:pdf_exact} is more concentrated around $\mathbb{E}\{T\}$, and the r.v. $T_k$ approaches a constant $T=1$.

\vspace{-5mm}

\begin{figure}[ht]
\includegraphics[width=0.85\linewidth]{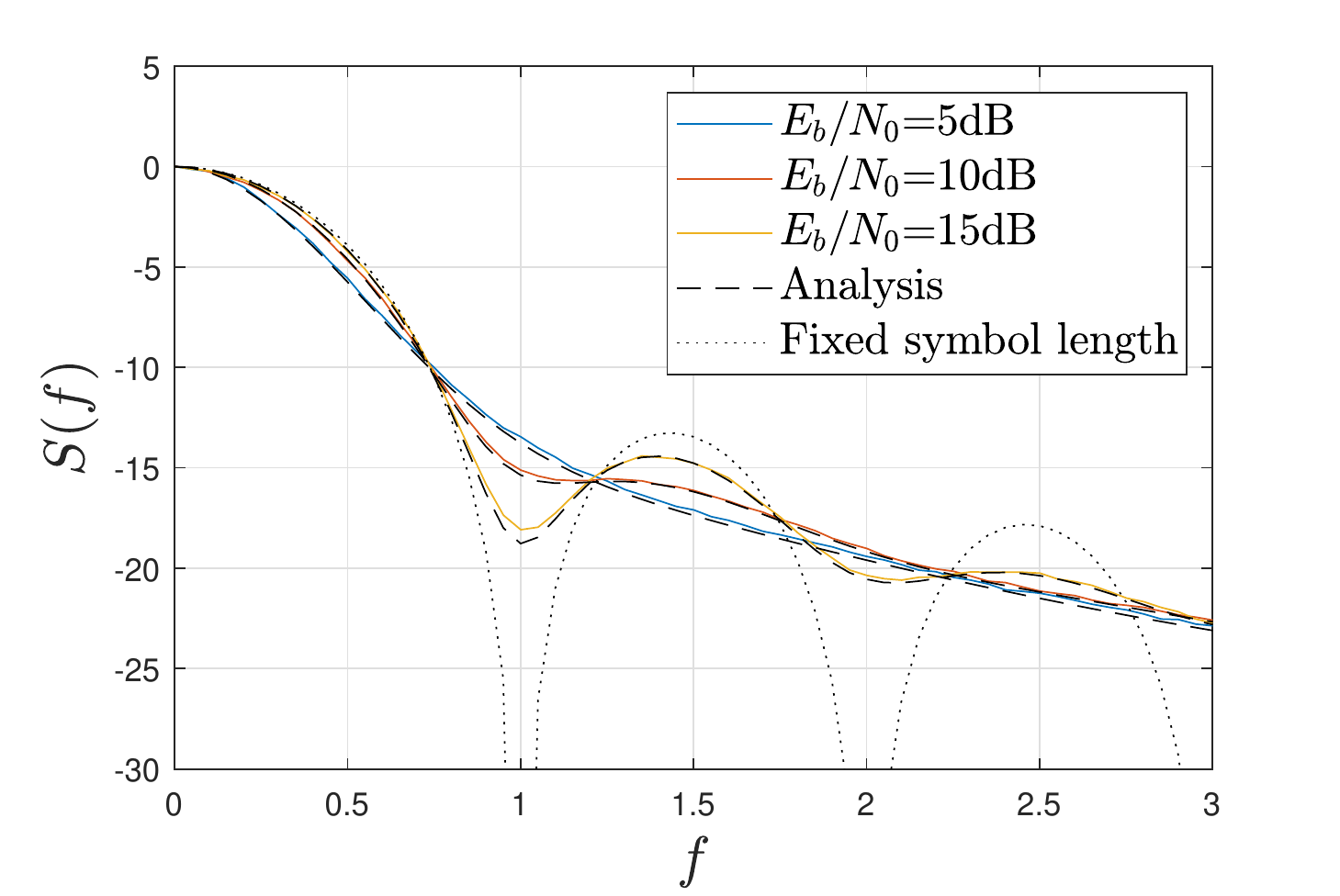}
\centering
\vspace{-2mm}
\caption{$S(f)$ of OSLA signal at various $E_b/N_0$.}
\label{fig:psd}
\end{figure}

\vspace{-2mm}
Finally, Figure \ref{fig:psd} shows the numerically evaluated PSD $S(f)$, which is the Fourier transform of $R(\tau)$.
From the figure, it is observed that the shape of spectrum changes with $\gamma=E_b/N_0$.
ITU-R defines the occupied bandwidth as the width of a frequency band such that, below the lower and above the upper frequency limits, the mean powers emitted are each equal to a specified percentage $\beta/2$ of the total mean power of a given emission \cite{ITUR_OBW}.
Under this definition with $\beta=0.05$ (i.e., the resulting bandwidth contains 95\% of the total power), the occupied bandwidth of the variable-length scheme with three different settings ($E_b/N_0=$5dB, 10dB, 15dB), and fixed-symbol-length BPSK is 1.89, 1.89, 1.85, and 1.74, respectively (for $\EE\{T\}=1$).

In practice, root-raised-cosine (RRC) pulse shaping is often used to reduce occupied bandwidth for a fixed-symbol-length scheme.
RRC pulse shaping is not directly applicable to OSLA due to the variable symbol length.
An alternative method to reducing the bandwidth of OSLA is using head/tail ramping up/down for each symbol. This may result in smoother pulses at the expense of reduced symbol rate. The analysis of such a scheme is non-trivial and its effects on the spectrum and error-rate performance is out of the scope of this paper.

\vspace{-4mm}

\vspace{-5mm}
\section{Conclusion}

In this letter, we provide autocorrelation and spectrum analysis for variable-symbol-length signals in communications with feedback.
The autocorrelation function $R(t,\tau)$ depends on both $t$ and $\tau$, however, it is observed that its value converges to a function that only depends on $\tau$ when $t \rightarrow \infty$ for an infinitely long sequence.
The spectrum can be numerically evaluated from the autocorrelation function $R(\tau )=\lim_{t \rightarrow \infty} R(t,\tau)$.
Applying the analysis results to a feedback-based communication scheme called OSLA, we confirm that the numerical results agree with the simulation.
Finally, we show that the variable-symbol-length scheme OSLA has the same spectral efficiency as a fixed-symbol length-scheme.

\vspace{-1mm}
\section*{Acknowledgement}
\small{This work was funded in part by NSF SWIFT Award \#2228974. }


\vspace{-1mm}
\bibliographystyle{IEEEtran}
\bibliography{IEEEabrv,ref}

\begin{thebibliography}{10}
\providecommand{\url}[1]{#1}
\csname url@samestyle\endcsname
\providecommand{\newblock}{\relax}
\providecommand{\bibinfo}[2]{#2}
\providecommand{\BIBentrySTDinterwordspacing}{\spaceskip=0pt\relax}
\providecommand{\BIBentryALTinterwordstretchfactor}{4}
\providecommand{\BIBentryALTinterwordspacing}{\spaceskip=\fontdimen2\font plus
\BIBentryALTinterwordstretchfactor\fontdimen3\font minus
  \fontdimen4\font\relax}
\providecommand{\BIBforeignlanguage}[2]{{%
\expandafter\ifx\csname l@#1\endcsname\relax
\typeout{** WARNING: IEEEtran.bst: No hyphenation pattern has been}%
\typeout{** loaded for the language `#1'. Using the pattern for}%
\typeout{** the default language instead.}%
\else
\language=\csname l@#1\endcsname
\fi
#2}}
\providecommand{\BIBdecl}{\relax}
\BIBdecl

\bibitem{Polyanskiy2011}
Y.~{Polyanskiy}, H.~V. {Poor}, and S.~{Verdu}, ``Feedback in the non-asymptotic
  regime,'' \emph{IEEE Transactions on Information Theory}, vol.~57, no.~8, pp.
  4903--4925, 2011.

\bibitem{Lott2007}
C.~Lott, O.~Milenkovic, and E.~Soljanin, ``Hybrid arq: Theory, state of the art
  and future directions,'' in \emph{2007 IEEE Information Theory Workshop on
  Information Theory for Wireless Networks}, 2007, pp. 1--5.

\bibitem{YI1979}
H.~{Yamamoto} and K.~{Itoh}, ``Asymptotic performance of a modified
  schalkwijk-barron scheme for channels with noiseless feedback (corresp.),''
  \emph{IEEE Transactions on Information Theory}, vol.~25, no.~6, pp. 729--733,
  1979.

\bibitem{Burnashev1976}
M.~V. {Burnashev}, ``{Data transmission over a discrete channel with feedback.
  Random transmission time},'' \emph{Problemy Peredachi Informatii}, vol.~12,
  no.~4, pp. 10--30, Oct.-Dec. 1976.

\bibitem{Viterbi1965}
A.~J. Viterbi, ``The effect of sequential decision feedback on communication
  over the gaussian channel,'' \emph{Information and Control}, vol.~8, no.~1,
  pp. 80--92, 1965.

\bibitem{OSLA_globecom}
C.-W. Hsu, A.~Anastasopoulos, and H.-S. Kim, ``Instantaneous feedback-based
  opportunistic symbol length adaptation for reliable communication,'' in
  \emph{2021 IEEE Global Communications Conference (GLOBECOM)}, 2021, pp.
  01--06.

\bibitem{deepcode2020}
H.~{Kim}, Y.~{Jiang}, S.~{Kannan}, S.~{Oh}, and P.~{Viswanath}, ``Deepcode:
  Feedback codes via deep learning,'' \emph{IEEE Journal on Selected Areas in
  Information Theory}, vol.~1, no.~1, pp. 194--206, 2020.

\bibitem{Ben2017}
A.~{Ben-Yishai} and O.~{Shayevitz}, ``Interactive schemes for the awgn channel
  with noisy feedback,'' \emph{IEEE Transactions on Information Theory},
  vol.~63, no.~4, pp. 2409--2427, 2017.

\bibitem{WienerBook}
A.~N. Borodin and P.~Salminen, \emph{Handbook of Brownian Motion - Facts and
  Formulae}.\hskip 1em plus 0.5em minus 0.4em\relax Birkhauser, 1996.

\bibitem{ITUR_OBW}
{ITU-R}, ``{Bandwidth measurement at monitoring stations [ITU-R SM.443-4]},''
  2007.

\end{thebibliography}

\end{document}